\documentclass[prl,twocolumn,showpacs,preprintnumbers,amsmath,amssymb]{revtex4}

\usepackage{graphicx}
\usepackage{dcolumn}
\usepackage{bm}

\begin{document}

\title{Vortex-antivortex nucleation in magnetically nanotextured superconductors: \\ Magnetic-field-driven and thermal scenarios}

\author{M. V. Milo\v{s}evi\'{c}}
\author{F. M. Peeters}
\email{Francois.Peeters@ua.ac.be}

\affiliation{Departement Fysica, Universiteit Antwerpen (Campus Middelheim), \\
Groenenborgerlaan 171, B-2020 Antwerpen, Belgium}

\date{\today}

\begin{abstract}
Within the Ginzburg-Landau formalism, we predict two novel
mechanisms of vortex-antivortex nucleation in a magnetically
nanostructured superconductor. Although counterintuitive,
nucleation of vortex-antivortex pairs can be activated in a
superconducting (SC) film covered by arrays of submicron
ferromagnets (FMs) when exposed to an external {\it homogeneous}
magnetic field. In another scenario, we predict the {\it thermal
induction} of vortex-antivortex configurations in SC/FM samples.
This phenomenon leads to a new type of Little-Parks oscillations
of the FM magnetization-temperature phase boundary of the
superconducting film.
\end{abstract}


\maketitle

Over the years, increased attention has been paid to experimental
and theoretical studies of vortex-antivortex (VAV) phenomena in
diverse systems. For example, the analogous process of
electron-positron pair creation and annihilation is well known in
quantum electrodynamics \cite{elpoz}. The appearance of VAV
structures has also been revealed in elastic waves in dusty
plasmas \cite{plasma}, Bose-Einstein condensates \cite{bec}, and
3D connected superfluid films \cite{3Dfluid}.

In the last decade, thermally activated VAV nucleation has been
extensively studied in two-dimensional superconducting (SC) films
\cite{gabay} and Josephson junctions \cite{fistul}. The
Berezinskii-Kosterlitz-Thouless (BKT) transition takes place at
finite temperature, below which the vortex-antivortex pairs are
bound and the sample resistance is zero \cite{kosterlitz}. In
another scenario, a VAV pair in SC film may be created (or
depinned) by a short lived local hot spot caused by the absorption
of a single photon \cite{kadin}. Recently, vortex-antivortex
structures were predicted in mesoscopic superconducting polygons
in homogeneous magnetic field \cite{chibotaru}. Those VAV states
are imposed by the symmetry of the sample, close to the nucleation
line $T_{c}(H)$. However, they become unstable deeper in the SC
state, except for effective type I superconductors \cite{kabanov}.

\begin{figure}[b]
\includegraphics[height=8cm]{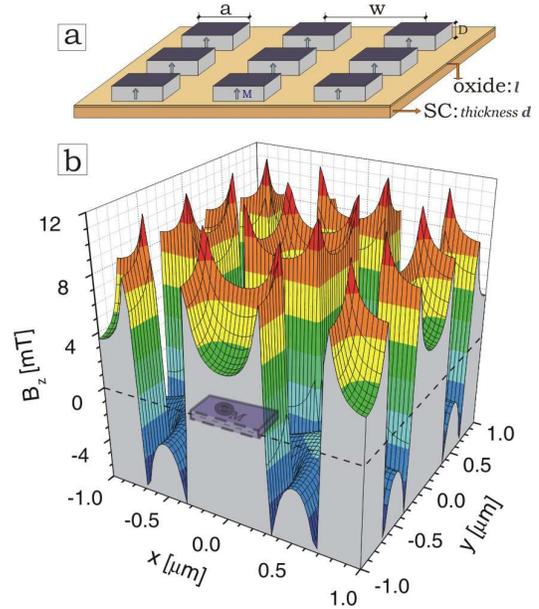}
\caption{\label{scheme} (a) A regular array of square magnetic
dots (MDs) on top of a superconducting (SC) film (separated by an
oxide layer); (b) the MDs-magnetic field profile in the SC plane
($a=0.6\mu$m, $W=1\mu$m, $D=25$nm, and bulk Co magnetization).}
\end{figure}

Intuitively, states containing vortex-antivortex pairs are
relevant to superconductors exposed to an {\it inhomogeneous}
magnetic field. In this respect, SC films with ferromagnetic (FM)
structures on top have been studied, and broken symmetry VAV
states \cite{priour}, VAV molecules and VAV lattices of different
geometries \cite{misko} have been predicted. The broader aspect of
the vortex-antivortex phenomena becomes apparent as the found
structures resemble the ones of electron dimples on the surface of
liquid helium, colloidal suspensions, dusty particles in complex
plasmas, and ionic or molecular crystals.

In this Letter, we report two new mechanisms of vortex-antivortex
nucleation in SC films under regular arrays of nanoengineered
FM-dots (see Fig. \ref{scheme}). Such hybrid structures have been
realized experimentally \cite{schuller, lange}. However, these and
following works were focused on commensurability effects with
magnetic pinning arrays, the consequent critical current
enhancement and the so-called field-induced superconductivity,
crucial for possible applications. The main objective of this
Letter is to explore the dynamical behavior of SC/FM samples when
exposed to {\it thermal excitations} and/or an additional {\it
homogeneous} magnetic field.

Our theoretical formalism relies upon the time-dependent
Ginzburg-Landau (GL) equations for the order parameter $\Psi$ and
the vector potential ${\bf A}$ \cite{kato}, supplemented by a
random force in order to include the thermodynamic fluctuations in
$\Psi({\bf r},t)$. In dimensionless form, when keeping the
temperature dependence explicitly, the GL equations become
\begin{equation}
\frac{\partial\Psi}{\partial t}=-\frac{1}{\zeta}\left[ \left(
-i\nabla-{\bf A}\right) ^{2}\Psi +\Psi \left(1-T-\left| \Psi
\right| ^{2}\right) \right]+ f({\bf r},t), \label{lijn1}
\end{equation}
\begin{equation}
\frac{\partial {\bf A}}{\partial t}=\textsf{Re}\left[\Psi ^{\ast
}\left(\frac{\bf \nabla}{i}-{\bf A}\right)\Psi\right]-\kappa
^{2}\nabla\times\nabla\times{\bf A}, \label{lijn2}
\end{equation}

\noindent where distances are expressed in units of $\xi(0)$, {\bf
A} in $H_{c2}(0)\xi(0)$, time in
$t_{0}=\pi\hbar\big/96k_{B}T_{c}$, temperature is scaled by
$T_{c}$, and $\Psi$ is normalized by its equilibrium value in the
absence of magnetic field. The relaxation constant $\zeta=12$ is
taken from the microscopic theory \cite{gork}. $f({\bf r},t)$ is
the dimensionless Langevin thermal noise, uncorrelated in space
and time $\langle f^{\ast}({\bf r},t)f({\bf
r}',t')\rangle=\frac{4\pi}{\zeta} E_{0}T\delta({\bf r}-{\bf
r}')\delta(t-t')$, where
$E_{0}=2k_{B}T_{c}\big/H_{c}(0)^{2}\xi(0)^{3}$ is the ratio of the
thermal energy to the free energy of a vortex. The random force
$f_{i}(t)$ at site $i$ is independently selected from a Gaussian
distribution with a zero mean.

For thin superconductors $(d<\xi ,\lambda)$, Eqs.
(\ref{lijn1},\ref{lijn2}) may be averaged over the SC thickness.
The lateral periodicity of the SC film (and the magnetic
structures on top of it) is included in our formalism through the
periodic boundary conditions ${\bf A}({\bf r}+{\bf b}_{i})={\bf
A}({\bf r})+{\nabla}\eta _{i}({\bf r})$, and $\Psi ({\bf r}+{\bf
b}_{i})=\Psi \exp(2\pi i\eta _{i}({\bf r})/\Phi _{0})$, where
${\bf b}_{i=x,y}$ are the periodicity vectors, and $\eta _{i}$ is
the gauge potential. The choice of gauge and details of the
numerical approach are explained in Ref. \cite{misko}. For given
magnetization of the ferromagnets $M$, we explore the behavior of
the superconducting state of the film starting the calculation
from randomly generated $(\psi,~{\bf A})$. In addition,
field-cooled and zero-field-cooled regimes are simulated when
Meissner state $(\Psi\approx 1)$ or the normal state $(\Psi\approx
0)$ are used as initial state. The ground state is then determined
by comparing Gibbs free energy ($\mathcal{G}$) of all found
states.

\begin{figure}[b]
\includegraphics[height=6.2cm]{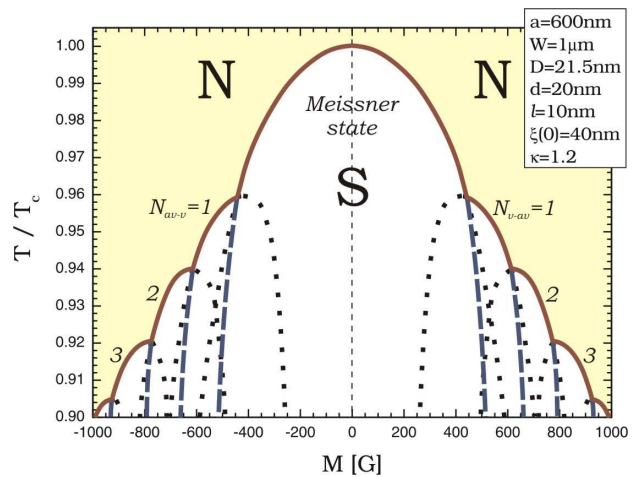}
\caption{\label{MT}Temperature versus MD-magnetization phase
diagram. The solid line is the superconducting/normal state
boundary. Dotted curves show the stability regions of different
vortex-antivortex configurations while dashed lines denote
transitions in the ground-state. For $M>0$, vortices are under the
MDs and antivortices between them (and vice versa). $N_{v-av}$
($N_{av-v}$) is the number of created pairs per MD.}
\end{figure}
{\it Influence of temperature.} The Little-Parks (LP) effect
\cite{lp}, i.e. oscillations of the critical temperature $T_{c}$
with change in the external magnetic field $H_{ext}$, is in
general associated with transitions between states with different
vorticity in multiply connected superconducting samples. In this
paper we focus on another possibility for creating multiquantum
vortex states: nucleation of the superconducting order parameter
in a hybrid system consisting of a thin superconducting film and
an array of square magnetic nanodots (MDs) as a function of
MD-magnetization $M$ (Fig. \ref{scheme}(a)). Provided that the
thickness of the SC film is small ($d<\xi$), the temperature of
the S/N transition as well as the structure of the superconducting
nuclei are mainly determined by a 2D distribution of the
MD-magnetic field (Fig. \ref{scheme}(b)). This magnetic field
landscape is inhomogeneous, with one crucial property - its total
flux through an infinite underlying plane equals {\it zero}. As a
result, vortices cannot nucleate without corresponding
antivortices, keeping the total vorticity at zero value.

\begin{figure}[t]
\includegraphics[height=6.2cm]{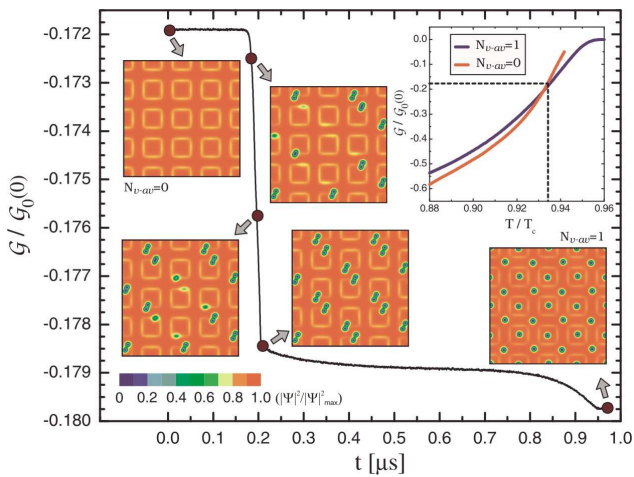}
\caption{\label{fluc}The $N_{v-av}$=0-1 transition at
$T/T_{c}$=0.935 for $M$=480G
($\mathcal{G}_{0}(0)$=$H_{c}(0)^2V\big/ 8\pi$, inset shows the
energy of the states as a function of temperature). The
$|\psi|^{2}$-density plots illustrate spontaneous
vortex-antivortex nucleation due to thermal fluctuations.}
\end{figure}

The calculated $M-T$ phase diagram is shown in Fig. \ref{MT} for
the typical values of the parameters corresponding to Pb or Nb
films. The solid line indicates the superconducting/normal state
(S/N) boundary ($|\psi|^{2}_{max}<10^{-5}$ denotes the normal
state). The dotted curves bound the stability regions of the
vortex-antivortex configurations deep in the SC state and dashed
lines denote the transitions in the ground-state. As one can see,
for example for $T/T_{c}=0.91$, increasing MD-magnetization leads
to the appearance of more vortex-antivortex pairs ($N_{v-av}$ is
the number of VAV pairs per MD), which arrange themselves in
lattices (see Ref. \cite{misko}).

The crucial result here is the LP-type of oscillations in the S/N
boundary. However, significant differences are present, as
compared to the usual LP-effect. Firstly, instead of the external
field, the magnetization of the MDs is changed. As a result, the
cusps in the phase boundary correspond now to the nucleation of
vortex-antivortex pairs, not to individual vortices. Secondly, the
total vorticity is always zero, and the analogue of the {\it
winding number} in our system is the number of (anti)vortices
(between) under the magnetic squares. For the same reason, the
position of cusps in the boundary will be determined by the
positive magnetic flux $\Phi^{+}$ under each magnet. However, flux
quantization does not have to be fulfilled, contrary to the
conventional LP-effect. Actually, $\Phi^{+}$ needed for the
appearance of the first VAV pair is larger than the quantized
$\Delta\Phi^{+}$ necessary for the nucleation of the following
pairs \cite{misko}. This LP-behavior of the S/N boundary with
variable period of the oscillations is the {\it first such
prediction for SC films}. The standard H$_{ext}$-T boundary for SC
films does not exhibit any oscillations \cite{lange}
(H$_{cr}$(T)=$\Phi_{0}\big/ 2\pi\xi$(T)$^{2}$). As explained in
Ref. \cite{misko}, vortex-antivortex configurations created by
regular MD-arrays are very compact and stable. Therefore, standard
transport measurements could confirm the predicted shape of the
phase boundary, through the measurements of the resistance
variations with changing magnetic moment of each MD (or
alternatively current in a loop) at a fixed temperature.

On the other hand, changing temperature for a fixed
MD-magnetization reveals another fascinating phenomenon. Namely,
in a zero-field cooled regime, the system relaxes in a vortex
state which is not necessarily the ground-state. If then
temperature is gradually changed (and thermal noise negligible),
the vorticity remains the same, due to finite energy barrier
(analogous to the Bean-Livingston one in finite samples), as long
as the given state is stable (see inset in Fig. \ref{fluc}).
However, increasing temperature strengthens thermal fluctuations
and lowers the barrier. As shown in Fig. \ref{fluc} for $M$=480G,
the initial $N_{v-av}$=0 state transits into $N_{v-av}$=1 as soon
as its energy becomes higher with increasing $T$. The energy
barrier at particular MDs can be overcome by the random thermal
fluctuations, and vortex-antivortex (VAV) pairs nucleate under the
MD-edge where current is maximal. Each VAV pair then causes VAV
nucleation at the neighboring site, due to the local supercurrent
enhancement (see next paragraph). Eventually, each MD bounds one
VAV pair and the system gradually relaxes into the ground-state
$N_{v-av}$=1 lattice. Therefore, vortex-antivortex pairs can be
induced in SC films {\it by increasing temperature}. Although the
nucleation of all VAV pairs takes less than $30$ns, the whole
$N_{v-av}=0\rightarrow 1$ transition takes a relatively long time
$t\approx 1\mu$s. This facilitates the experimental observation of
the thermally induced VAV pairs, by e.g. Scanning Hall Probe,
Magnetic Force, or Lorentz Microscopy.

\begin{figure}[b]
\includegraphics[height=6.3cm]{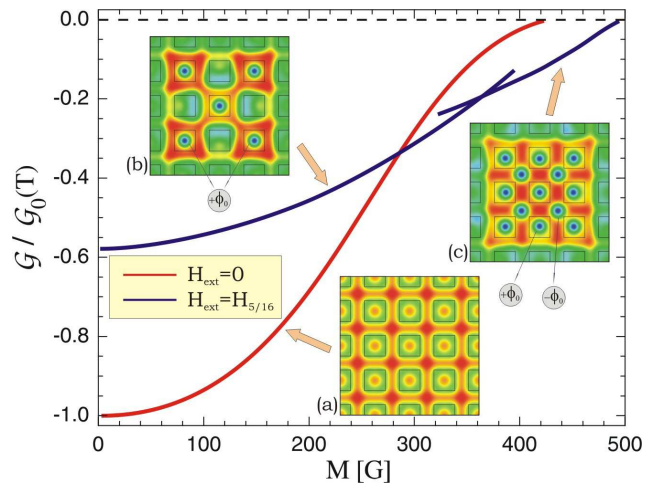}
\caption{\label{freeVAV}The Gibbs free energy as a function of
MD-magnetization at $T/T_{c}=0.97$, in the case of no applied
homogeneous magnetic field $H_{ext}=0$ and rational
$H_{ext}=H_{5/16}$ field. The insets show the Cooper-pair density
plots: (a) Meissner state, (b) 5 pinned vortices, and (c)
vortex-antivortex configuration with total vorticity 5.}
\end{figure}
{\it Influence of applied homogeneous field.} If our sample is
exposed to an additional homogeneous magnetic field $H_{ext}$, we
expect the symmetry breaking of the phase boundary of Fig.
\ref{MT}, due to the interaction between the MD-induced VAV pairs
and external flux lines. Namely, MDs attract the added vortices
for ${\bf M}\|\textbf{H}_{ext}$, and vice versa \cite{schuller}.
In what follows, we will focus on the part of the M-T diagram
where no MD-induced VAV pairs exist, namely for $1>T/T_{c}>0.96$.
At these temperatures, increasing $M$ continuously suppresses the
order parameter under MDs (see Fig. \ref{freeVAV}, inset (a)),
until the transition to the normal state (red curve in Fig.
\ref{freeVAV}). Although critical conditions are met, no
vortex-antivortex pairs can be induced in the underlying
superconductor, even in the case of very strong magnets. Namely,
vortices and antivortices cannot be adequately separated and
stabilized since magnetic lattice is too dense compared to
$\xi$(T) \cite{misko}.

\begin{figure}[t]
\includegraphics[height=8.3cm]{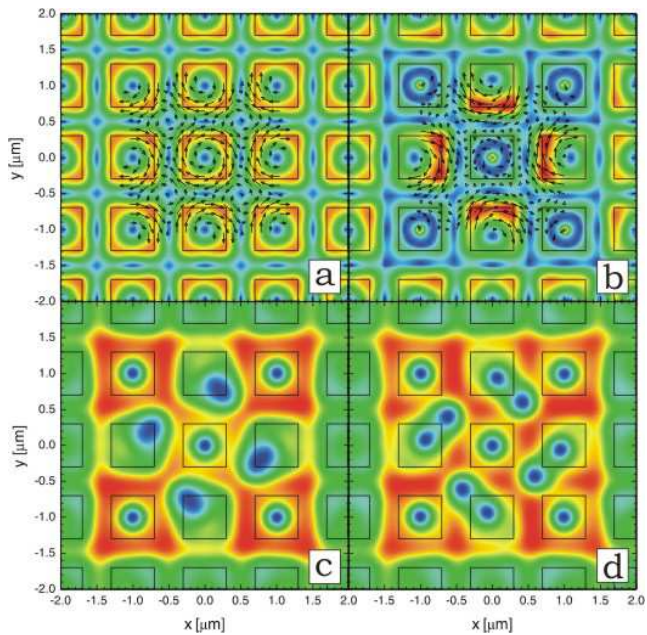}
\caption{\label{curvav} The superfluid velocity contourplots
($v_{max}=v_{cr}$) and the corresponding vectorplots for $M$=400G
(see Fig. \ref{freeVAV}) and: (a) no applied field, (b) $H_{ext}$=
0.646mT, just before the VAV nucleation. (c-d) the snapshots of
the subsequent VAV nucleation process ($|\psi|^{2}$
contourplots).}
\end{figure}
However, the very same SC/FM sample demonstrates different
behavior when exposed to homogeneous magnetic field
($\textbf{H}_{ext}\|{\bf M}$). We applied $H_{ext}=0.646$mT,
creating five flux quanta piercing through the 4x4 unit cell. For
weaker MDs, conventional pinning phenomena are observed, as
external vortices are pinned by MDs (see \cite{schuller}), in a
specific regular configuration (see Fig. \ref{freeVAV}, inset
(b)). However, for higher magnetization, two fascinating phenomena
occur: (i) the energy of the sample exposed to $H_{ext}$ becomes
{\it lower} than the one in $H_{ext}$=0 case, and (ii) VAV pairs
nucleate and the most favorable vortex state consists of 9
vortices under MDs interconnected by 4 interstitial antivortices
(see Fig. \ref{freeVAV}, inset (c)).

The explanation for these findings lies in the {\it
supercurrents}. The vortex nucleation in superconductors is
governed by the supervelocity $\textbf{\textit{v}}\sim
\textbf{\textit{j}}\big/ |\psi|^{2}$, plotted in Fig.
\ref{curvav}(a-b), for $M$=400G without and with applied
homogeneous field. For plain SCs in external homogeneous field,
the vortex nucleation occurs for ${\it v}_{cr}\approx 1$, but for
a VAV pair induced by a magnet this value is roughly $50\%$ larger
[note the non-quantized $M_{cr}$ (i.e. $\Phi^{+}$) in Fig.
\ref{MT}] and depends on the magnet parameters. The
$\textbf{\textit{v}}$ vectorplot in Fig. \ref{curvav}(a) shows
that MDs induce antivortex-like supercurrents, which compensate
each other in interstitial areas and are maximal under the
MD-edges. As explained above, for $H_{ext}=0.646$mT, bringing 5
flux quanta in the region depicted in Fig. \ref{curvav}, added
vortices are pinned by MDs in a regular formation. Therefore, at
the pinning sites, MD-induced currents are compensated by the
vortex currents (see Fig. \ref{curvav}(b)), which lowers the total
energy and explains phenomenon (i). This current compensation also
causes enhanced superconductivity in SC/FM samples when exposed to
additional field (e.g. higher $M_{cr}$ in Fig. \ref{freeVAV}),
contrary to conventional behavior, and explains the recent
experimental findings of field-induced superconductivity
\cite{lange}. On the other hand, the currents under MDs
neighboring to the pinning sites are {\it enhanced}. As shown in
(b), critical supervelocity ${\it v}_{cr}$=1.447$H_{c2}\xi$ is
reached, and VAV pairs may nucleate. More importantly, the
compensation of currents at the pinning sites leads to larger
effective interstitial space for the nucleated antivortices.
Therefore, contrary to the case in the absence of $H_{ext}$, there
are no obstacles for VAV nucleation, and Figs. \ref{curvav}(c-d)
show snapshots of that process, on the path to the final
configuration shown as inset (c) in Fig. \ref{freeVAV}.

In conclusion, we described two novel scenarios of
vortex-antivortex (VAV) nucleation in a SC film covered by
magnetic dot (MD) arrays. The inhomogeneous stray field of the
dots stimulates the appearance of VAV pairs. However, even when
conditions for VAV nucleation are not fulfilled, the creation of
these fascinating structures can be driven either by temperature
variations, or applying a {\it homogeneous} magnetic field.
Changing the temperature leads to a Little-Parks-like M-T phase
boundary, where each cusp corresponds to the creation of a new
VAV-pair per MD. Although counterintuitive, the applied
homogeneous magnetic field embraces the creation of VAV-pairs, in
cases when proximity of the magnets (compared to $\xi$(T)) does
not support it. Pinned vortices at given MDs contribute to the
critical nucleation conditions under the neighboring MDs in the
lattice through the supercurrent compensation. As a result, a new
type of VAV lattice is formed, with positive net vorticity.

The authors acknowledge support from the Flemish Science
Foundation (FWO-Vl), the Belgian Science Policy, and the ESF
programme ``VORTEX''.

\end{document}